\documentclass[prl,twocolumn,twoside,aps,floatfix,longbibliography,natbib,nofootinbib,superscriptaddress]{revtex4-1}

\usepackage{amsmath, bm, amsfonts, amssymb}     
\usepackage{graphicx}                       
\usepackage{xfrac}                          
\usepackage{grffile}                        
\usepackage{color,colortbl}
\usepackage{xcolor}

\usepackage[colorlinks, citecolor=blue, urlcolor=red, linkcolor=blue]{hyperref}

\renewcommand{\eqref}{Eq.~\ref}

\newcommand{\gdot}[0] {\dot{\gamma}}

\newcommand{\gdotbar}[0] {\bar{\dot{\gamma}}}

\newcommand{\xhat}{\hat{\mathbf{x}}}
\newcommand{\yhat}{\hat{\mathbf{y}}}

\newcommand{\tens}[1]{\mathbf{{#1}}}

\newcommand{\tw}{t_{\rm w}}

\newcommand{\be}{\begin{equation}}
\newcommand{\ee}{\end{equation}}
\newcommand{\bea}{\begin{eqnarray}}
\newcommand{\eea}{\end{eqnarray}}

\newcommand{\smf}[1] {\textcolor{black}{#1}}

\begin{document}

\title{Ultra-delayed material failure via shear banding after straining an amorphous material}

\author{Henry A. Lockwood}
\author{Emily S. Carrington}
\author{Suzanne M. Fielding}

\affiliation{Department of Physics, Durham University, Science Laboratories,
  South Road, Durham DH1 3LE, United Kingdom}

\begin{abstract}
We predict a phenomenon of catastrophic material failure arising suddenly within an amorphous material, with an extremely long delay time since the material was last deformed. By simulating a mesoscopic soft glassy rheology model in one dimension (1D), a mesoscopic elastoplastic model in 1D and 2D, and a  continuum fluidity model in 1D, we demonstrate the basic physics to involve a dramatic ultra-delayed shear banding instability, in which strain suddenly strongly localises within the material and the stress drops precipitously. The delay time after the long historical shear strain was applied before failure occurs increases steeply with decreasing strain amplitude, decreasing working temperature, and increasing sample annealing prior to shear. In demonstrating the same physics -- which is directly testable experimentally and in particle simulations --- to obtain within three different constitutive models, we suggest it may be generic across amorphous materials. The counter-intuitive prediction of catastrophic  material failure long after any deformation was last applied could have important consequences for material processing and performance.
\end{abstract}

\maketitle

Amorphous materials~\cite{nicolas2018deformation} include soft glasses~\cite{sollich1997rheology} and yield stress
fluids~\cite{bonn2017yield}   such as dense emulsions, foams, colloids, gels, and granular matter; as well as harder metallic and molecular glasses~\cite{hufnagel2016deformation,greer2013shear}. Unlike in conventional crystalline materials, the internal mesoscopic substructures of which amorphous materials are formed -- emulsion droplets, colloidal particles, sand grains, {\it etc.} -- lack long-ranged crystalline order. In consequence, understanding the rheological (deformation and flow) properties of such materials presents a considerable challenge.

When subject to small loads, amorphous materials typically show an elastic solid-like response. At larger loads,
they exhibit a dynamical plastic yielding transition. Important practical processes governed by dynamical yielding include the rising of bubbles in radioactive sludge~\cite{gens2009full1}; 
cement spreading~\cite{chidiac2009plastic,banfill1981viscometric1};  the yielding of food being chewed~\cite{fischer2011rheology1}; the
material failure of hard glasses~\cite{greer2013shear}; geophysical processes such as lava flows, earthquakes and mudslides ~\cite{mader2013rheology1,daub2010pulse1,coussot2002avalanche} and the breakup of sea-ice~\cite{feltham2008sea}; and the reshaping of biological tissue~\cite{ranft2010fluidization1,gonzalez2012soft1,park2015unjamming1}.

Experimentally, a common protocol consists of subjecting an amorphous material at some time $t=0$ to the switch-on of a shear strain of some rate $\gdot$, which is held constant for all $t>0$. Typically, after an initially elastic response in which the shear stress $\Sigma$ rises roughly linearly with the accumulating strain $\gamma=\gdot t$, the stress attains a maximum then declines as yielding initiates.  Alternatively, a material can be subject to the switch-on at $t=0$  of a shear stress of amplitude $\Sigma$. A typical response then comprises a lengthy regime of initial slow creep, in which the strain rate decays as $\gdot(t)\sim t^{-\alpha}$, with $0<\alpha<1$, and the strain increases sub-linearly, $\gamma(t)\sim t^{1-\alpha}$, before the sample suddenly yields and the strain rate increases dramatically~\cite{miguel2002dislocation,bauer2006collective,gibaud2010heterogeneous,chaudhuri2013onset,leocmach2014creep,sentjabrskaja2015creep,ballesta2016creep,landrum2016delayed,liu2018creep,cabriolu2019precursors,siebenburger2012creep}. 

In both these experiments, {\em shear localisation} is commonly observed as yielding  initiates~\cite{divoux2010transient,martin2012transient,gibaud2008influence,dimitriou2014comprehensive,colombo2014stress,shi2007evaluation,shrivastav2016heterogeneous,fielding2014shear,moorcroft2011age,manning2007strain1,manning2009rate1,hinkle2016small1,jagla2010shear1,divoux2011stress1,divoux2012yielding1,grenard2014timescales,gibaud2009shear,gibaud2010heterogeneous,gibaud2008influence,kurokawa2015avalanche,greer2013shear,divoux2011stress1,divoux2010transient,divoux2012yielding1,grenard2014timescales,gibaud2009shear,gibaud2010heterogeneous,gibaud2008influence,kurokawa2015avalanche,sentjabrskaja2015creep,koumakis2016start,ballesta2016creep,mohraz2005orientation,koumakis2012yielding,rogers2008aging}, often with a state of initially homogeneous shear in the early time solid-like regime giving way to forming shear bands of differing strain rate, with layer normals in the shear gradient direction. Indeed, it can be shown within a minimal set of continuum rheological constitutive assumptions that an initially homogeneous shear often becomes linearly unstable to shear banding as yielding initiates~\cite{fielding2016triggers,moorcroft2014shear,moorcroft2013criteria,fielding2013modeling,adams2011transient,moorcroft2011age,fielding2014shear}. 

To date, most studies have focused on  experiments of the kind just described,  in which the shear strain $\gamma$ accumulates indefinitely in an ongoing way (post switch-on),  whether under a sustained imposed shear rate $\gdot$ or shear stress $\Sigma$, and the material yields into a final steady flowing state~\cite{divoux2010transient,shrivastav2016heterogeneous,vasisht2020computational,vasisht2020emergence,shrivastav2016yielding,alix2018shear,divoux2012yielding1,grenard2014timescales,benzi2019unified,chaudhuri2013onset,manning2007strain1,hinkle2016small1,jagla2007strain} (or fails catastrophically en route
~\cite{hufnagel2016deformation,chen2008mechanical,boyce1988large1,anand2012large1,ward2012mechanical1,doyle1972fracture1,huang2016multiple,baljon1996energy,bonn1998delayed,ducrot2014toughening,tabuteau2009microscopic,leocmach2014creep,hufnagel2016deformation,chen2008mechanical,boyce1988large1,anand2012large1,ward2012mechanical1}). Others consider yielding in cyclically repeating forward and reverse straining that likewise continues indefinitely over time~\cite{perge2014time,saint2017predicting,gibaud2016multiple,LAOS_ARXIV}.

Here we consider a simpler scenario, in which an amorphous material is instead subject to a {\em finite} shear strain $\gamma_0$, applied once at time $t=0$ and held constant thereafter, with no further deformation applied. Given the absence of any finally flowing state or repeating strain in such a scenario, it
has been implicitly assumed that the stress relaxation post-strain will take place in a straightforwardly innocuous way as the material slowly recovers. 

On the contrary, our central contribution will be to show that an amorphous material can instead suffer a dramatic internal shear banding instability, in which the strain field  suddenly becomes highly heterogeneous across it, leading to  a precipitous stress drop. Crucially, this catastrophic material failure can be delayed an arbitrarily long time $t^*$ after the strain was applied, with $t^*$ increasing dramatically with decreasing amplitude of the imposed strain $\gamma_0$, decreasing working temperature $T$ and increasing degree of sample annealing prior to shear. This prediction of catastrophic material failure arising a potentially indefinitely long time after a material was last (externally) deformed is highly counter-intuitive. Indeed, an observer lacking knowledge of the strain history could be caught unawares by the instability, with important consequences for material processing and performance.

 In substantiating this  basic physics within three different constitutive models -- a mesoscopic soft glassy rheology (SGR)~\cite{sollich1997rheology}, a mesoscopic elastoplastic model (EPM)~\cite{nicolas2018deformation}, and a  continuum fluidity model~\cite{moorcroft2011age} -- we suggest it may be generic across amorphous materials.

\section*{Theoretical models}

We consider a slab of material sandwiched between infinite flat parallel plates at $y=0, L_y$.  At time $t=0$ an initially uniform shear strain of amplitude $\gamma_0$ is imposed, giving a displacement $\textbf{u}(\textbf{r},t)=u(y)\xhat=\gamma_0 y\xhat$ along $\xhat$, with shear gradient direction $\yhat$.
No further (global) strain is imposed, so the average shear rate across the sample
$\gdotbar
\equiv
\int_0^{L_y}\gdot(y,t)dy =0$ for any time $t>0$.
If shear banding later arises, however,  it causes internal deformations within the material and the strain field becomes heterogeneous, subject to the displacement and velocity fields $\textbf{u}(\textbf{r},t)$ and $\textbf{v}(\textbf{r},t)$ obeying the incompressibility condition, $\nabla.\textbf{u}=\nabla.\textbf{v}=0$.  In most of what follows we adopt a 1D approach, in which displacements still arise only along $\xhat$  and gradients only along $\yhat$, such that $\textbf{u}=u(y,t)\xhat$,  with strain field $\gamma(y,t)=\partial_y u(y,t)$, velocity $v(y,t)=\dot{u}(y,t)$ and strain-rate $\gdot(y,t)=\partial_y v(y,t)$. 

In any material element, we assume the total stress
$\mathbf{\Sigma}=
\boldsymbol{\sigma} + 2\eta\tens{D} -p\tens{I}$
to comprise an elastoplastic contribution $\boldsymbol{\sigma}$ from the mesoscopic substructures (emulsions droplets,  etc), a solvent contribution of viscosity $\eta$, and a pressure, $p$. Here $\tens{D}=\tfrac{1}{2}(\tens{K} +\tens{K}^T)$ with $K_{\alpha\beta} = \partial_\beta v_\alpha$.  
 The stress field 
$\mathbf{\Sigma}(\textbf{r},t)$ obeys the 
force balance condition,
$\nabla.\mathbf{\Sigma} = 0$. In 1D we track only the shear stress $\Sigma_{xy}=\sigma_{xy}+\eta\gdot$, and drop the $xy$ subscript.

To model the dynamics of the elastoplastic stress $\sigma$, the SGR model~\cite{sollich1997rheology} considers an ensemble of elements, each corresponding to a mesoscopic region of material.  Under an imposed shear rate $\gdot$, each element experiences a buildup of local elastic shear strain $l$ with $\dot{l}=\gdot$, and stress $Gl$, with modulus $G$. This is
intermittently released by local plastic yielding events, modelled as hopping of an element over a strain-modulated energy barrier $E$, governed by a temperature $T$, with
yielding intervals chosen stochastically with rate
$\tau_0^{-1}{\rm max}\left\{\exp[-(E-\tfrac{1}{2}kl^2)/T],1\right\}$. 
Upon yielding, an element resets its local stress to zero and selects its new energy barrier  from a distribution $\rho(E)\sim \exp(-E/T_{\rm g})$.   This  confers a broad spectrum of yielding times and a glass phase for $T<T_{\rm g}$, in which the timescale for relaxation of the macroscopic stress $\sigma=G\langle l\rangle$ following a small step strain increases linearly with  sample age $\tw$~\cite{derec2003aging1,fielding2000aging1}.  

The formation of 1D shear bands with layer normals along $\yhat$ is accounted for by discretising $y=0...L_y$ into $s=1...S$ streamlines, with
periodic boundary conditions~\cite{fielding2009shear1}. On each streamline are placed $m=1...M$ SGR elements.  The elastoplastic stress  on streamline $s$ is $\sigma_s = (G/M)\sum_m l_{sm}$.  Given an imposed average shear rate $\gdotbar$ across the sample as a whole, the streamline shear rate is calculated by enforcing force balance, as $\gdot_s=\gdotbar+(\bar\sigma-\sigma_s)/\eta$, where $\bar\sigma=(1/S)\sum_s\sigma_s$. 

Our 1D EPM is defined likewise, except that each element now has the same yield energy $E$, and after yielding selects its new local strain $l$ from a Gaussian  of small width $l_{\rm h}$ To solve this model numerically across $S$ streamlines, we evolve the Fokker-Planck equation for the  distribution of elemental strains on each streamline, with the streamline strain rate calculated by enforcing force balance (as in SGR). The distribution on each streamline is discretized on a grid of $Q$ values of $l$, linearly space between $l=-l_{\rm c}$ and $l=+l_{\rm c}$, chosen large enough to encompass the full distribution. To confirm that the phenomenon we report arises beyond 1D, we also present results for the same EPM in 2D~\cite{nicolas2018deformation}, with one elastoplastic element on each of $N^2$ lattice sites and Eshelby stress propagation implemented instantaneously after local yielding as in~\cite{pollard2022yielding}.

The fluidity model supposes a Maxwell-type constitutive
equation for the elastoplastic stress
\be
\partial_t\sigma(y,t)=G\gdot-\sigma/\tau,
\label{eqn:sigma}
\ee
where $G$ is a constant modulus and $\tau$ is a structural relaxation
time (inverse fluidity) that has its own dynamics:
\be
\partial_t\tau=f(\tau,\sigma,\gdot)+l_o^2\partial^2_y \tau.
\label{eqn:tau}
\ee
In this equation, $f=1-|\gdot|(\tau-\tau_0)(|\sigma|-\sigma_{\rm
th})\Theta(|\sigma|-\sigma_{\rm th})$, with $\sigma_{\rm th}=1$. Like SGR, this model captures rheological aging, with
the timescale for stress relaxation following the imposition of a step
strain increasing linearly with the system age, $\tau =\tw$. A steady
flow cuts off ageing at the inverse strain rate, and the steady state
flow curve displays a yield stress.   The parameter $l_o$ in Eqn.~\ref{eqn:tau} is a mesoscopic
length describing the tendency for the relaxation time of a
mesoscopic region to equalise with those of its neighbours.


We choose units $G=\tau_0=L_y=1$ and rescale strain such that $T_{\rm g}=1$ (SGR) and  $E=1$ (EPM). The imposed strain $\gamma_0$ is thus scaled by the typical local yield strain, which varies between experimental systems, and does not represent an absolute strain. For the solvent viscosity $\eta\ll G\tau_0=1$ we set $\eta=0.05$ in the SGR and fluidity models, $\eta=0.01$ in 1D EPM, and $\eta=0$ in 2D EPM.   We consider $S=10$ streamlines in the 1D EPM and SGR models, and the continuum limit $S\to\infty$ in the fluidity model. Our findings will prove robust to these variations in $\eta$ and $S$. The numerical timestep $\Delta t \to 0$. In SGR we take $M=100,000$ elements per streamline. In EPM we consider a
grid of $Q=100, 000$ values of $l$, linearly space between $l=-l_{\rm c}$ and $l=+l_{\rm c}$, with $l_{\rm c}=10.0$

\subsection*{Sample preparation}

The importance to yielding of sample preparation, annealing and initial disorder is increasingly being
appreciated~\cite{leishangthem2017yielding1,shi2005strain,popovic2018elastoplastic1,barlow2020ductile1,vasisht2020emergence}. Accordingly, we shall model two different sample preparation protocols. Within the SGR and fluidity models, we prepare the system via a  quench at time $t=-\tw$, from an infinite temperature to a working temperature $T<T_{\rm g}$, then allow it age for a waiting time $\tw$ before applying the shear at $t=0$.
Within EPM, we instead  equilibrate the sample to a temperature $T_0$, giving a Gaussian distribution of local strains of variance $T_0$, then suddenly at time $t=0$ quench to a working temperature $T<T_0$, then immediately apply the strain.  A larger  $\tw$ (SGR and fluidity) or smaller $T_0$ (EPM) corresponds to a better annealed sample. 

About an initially uniform shear state, small random heterogeneity is seeded naturally via $M$ and $S$ being finite. We also seed small systematic perturbations as   $\tw \to \tw \left( 1 + \epsilon\cos{2\pi y}\right)$ with $\epsilon=0.1$ in the SGR and fluidity models, and $\gamma\to\gamma \left( 1 + \delta\cos{2\pi y}\right)$ with $\delta=0.05$ in 1D EPM. 

Our findings below will prove robust to these different methods of sample preparation and seeding.


%
\begin{figure}[!t]
  \includegraphics[width=8.5cm]{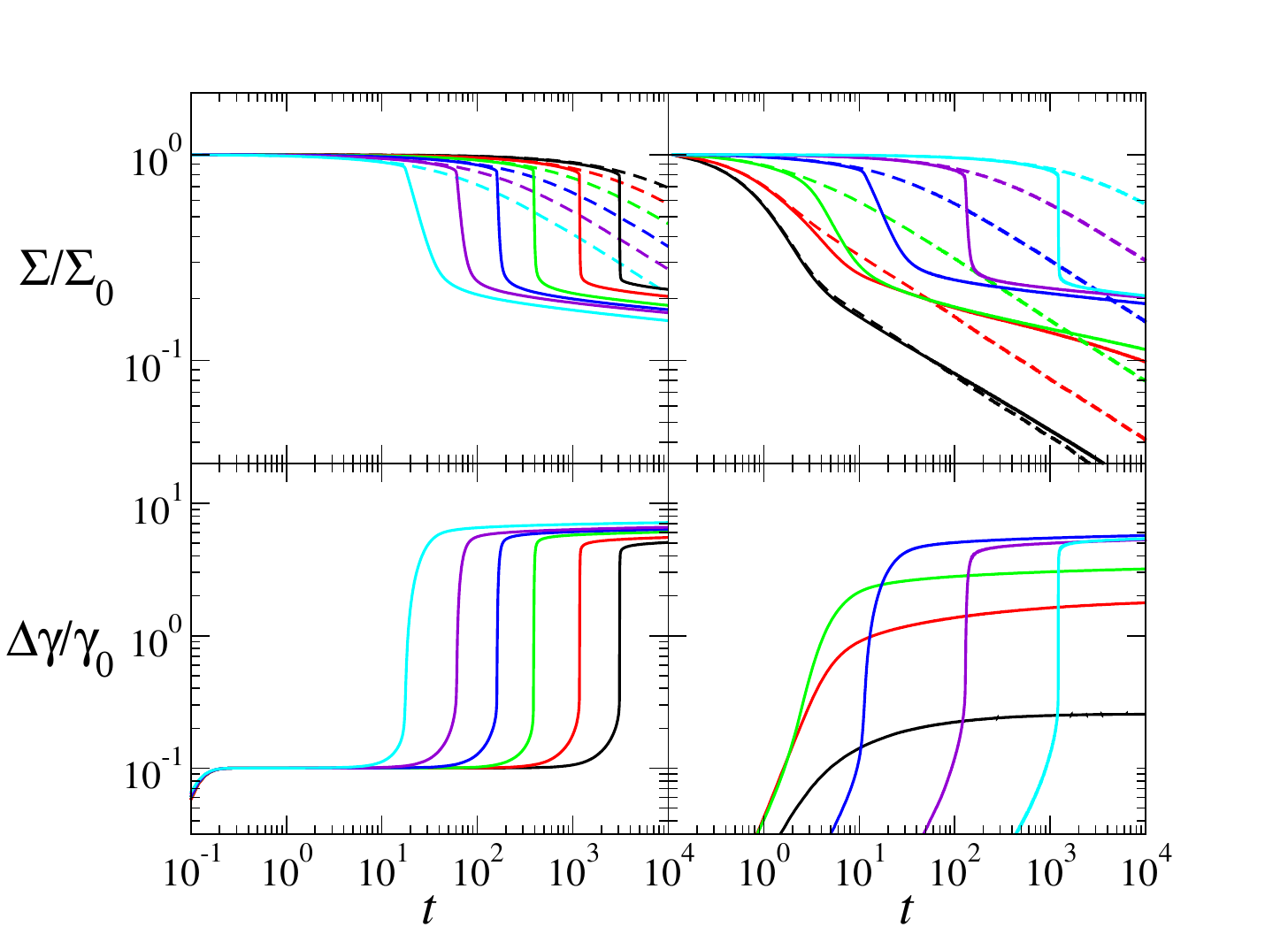} 
  \caption{{\bf Soft glass rheology model.}  {\bf Top:} Stress
  decay over time $t$ since the
  imposition of a step shear strain. Dashed curves show calculations in which
  the strain field is artificially constrained to remain homogeneous,
  solid in which it is allowed to become heterogeneous. {\bf Bottom:} Corresponding degree of strain heterogeneity $\Delta\gamma$, normalised by the imposed strain $\gamma_0$. {\bf Left:} Strain amplitude $\gamma_0=2.4, 2.5\cdots 2.9$ in curve sets right to left, waiting time  $\tw=10^8$.
  {\bf Right:} Strain amplitude $\gamma_0=2.5$, waiting
  times  $\tw=10^3,10^4,\cdots 10^8$ in curve sets left to right.
  Working temperature $T=0.3$.   }
   \label{fig:SGRmodel_time_new}
\end{figure}

\section*{Results}

\subsection*{Soft glassy rheology model}

The basic physics that we report is demonstrated in the SGR model in Fig.~\ref{fig:SGRmodel_time_new}. The top left panel shows the stress versus time $t$ following the imposition of a step shear strain at $t=0$, for several strain amplitudes $\gamma_0$, at fixed sample annealing  prior to shear, $\tw$, and  working temperature $T$. (Each curve corresponds to a single numerical run, with noise irrelevant.) The solid lines show  calculations in which the strain field $\gamma(y,t)$ is allowed to  become heterogeneous across the gradient direction $y$; the dashed lines in which it  is artificially constrained to remain homogeneous, $\gamma(y)=\gamma_0$, independent of $y$. The departure of the former from the latter marks the onset of a dramatic shear banding instability in which the strain $\gamma(y,t)$  becomes highly heterogeneous. To characterise this heterogeneity, we define $\Delta\gamma(t)$  at any time as the maximum minus the minimum of $\gamma(y,t)$ across  $y$.  This suddenly  increases from a small initial value to a large value $O(10\gamma_0)$ (bottom left panel) as the stress drops precipitously (top left). 

\begin{figure}[!t]
  \includegraphics[width=8.5cm]{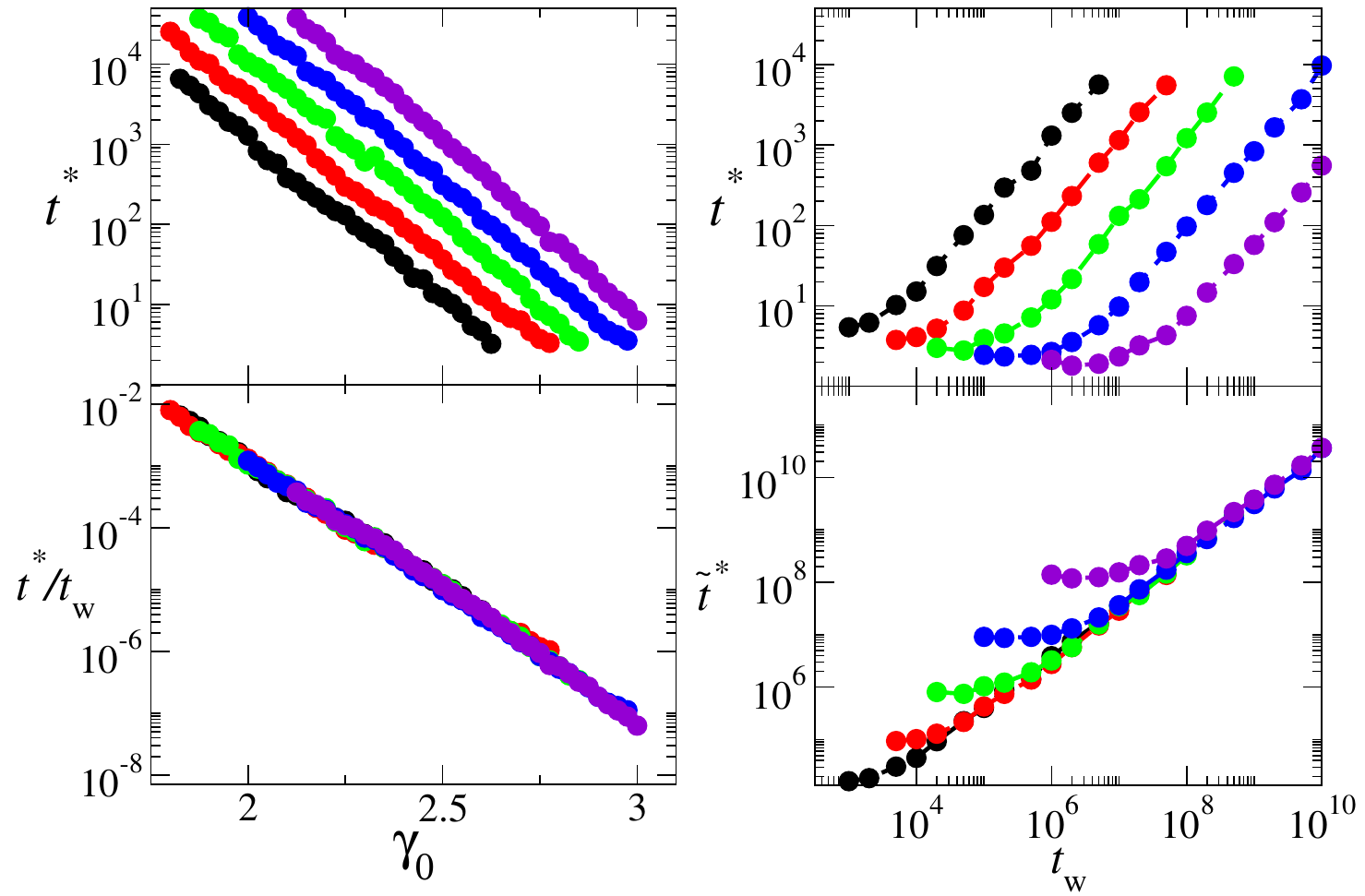}
  \caption{{\bf Soft glassy rheology model.} Time delay $t^*$ after step strain before material failure arises.   {\bf Top left:} raw $t^*$ versus strain amplitude $\gamma_0$ for waiting times $\tw=10^n$ with $n=6.0, 6.5\cdots 8.0$ in data sets upwards. {\bf Bottom left:} same data scaled by the waiting time $\tw$. {\bf Top right:} raw $t^*$ versus sample age $\tw$ at the time the strain is imposed, for strain amplitudes $\gamma_0= 2.00, 2.25,\cdots 3.00$ in curves  left to right. {\bf Bottom right:} same data scaled as $\tilde{t^*}=t^* \exp(\alpha\gamma_0^2/2x)$,  $\alpha\approx 1.2$. Working temperature $T=0.3$.}
  \label{fig:SGR_tStar_vs_tw_and_gamma0}
\end{figure}

Remarkably, this phenomenon of catastrophic  material failure can arise a very long time after the strain was imposed, with a delay time $t^*$ that increases significantly with decreasing imposed strain amplitude.  To quantify this, we define  $t^*$ as the time at which  $\Delta \gamma(t)$ grows most quickly with time. This quantity is plotted as a function of strain amplitude in Fig.~\ref{fig:SGR_tStar_vs_tw_and_gamma0} (top left), for several waiting times $\tw$. Delay times $t^*>10^4$ at even lower strain amplitudes are unfeasible to access numerically.

\begin{figure}[!t]
  \includegraphics[width=8.0cm]{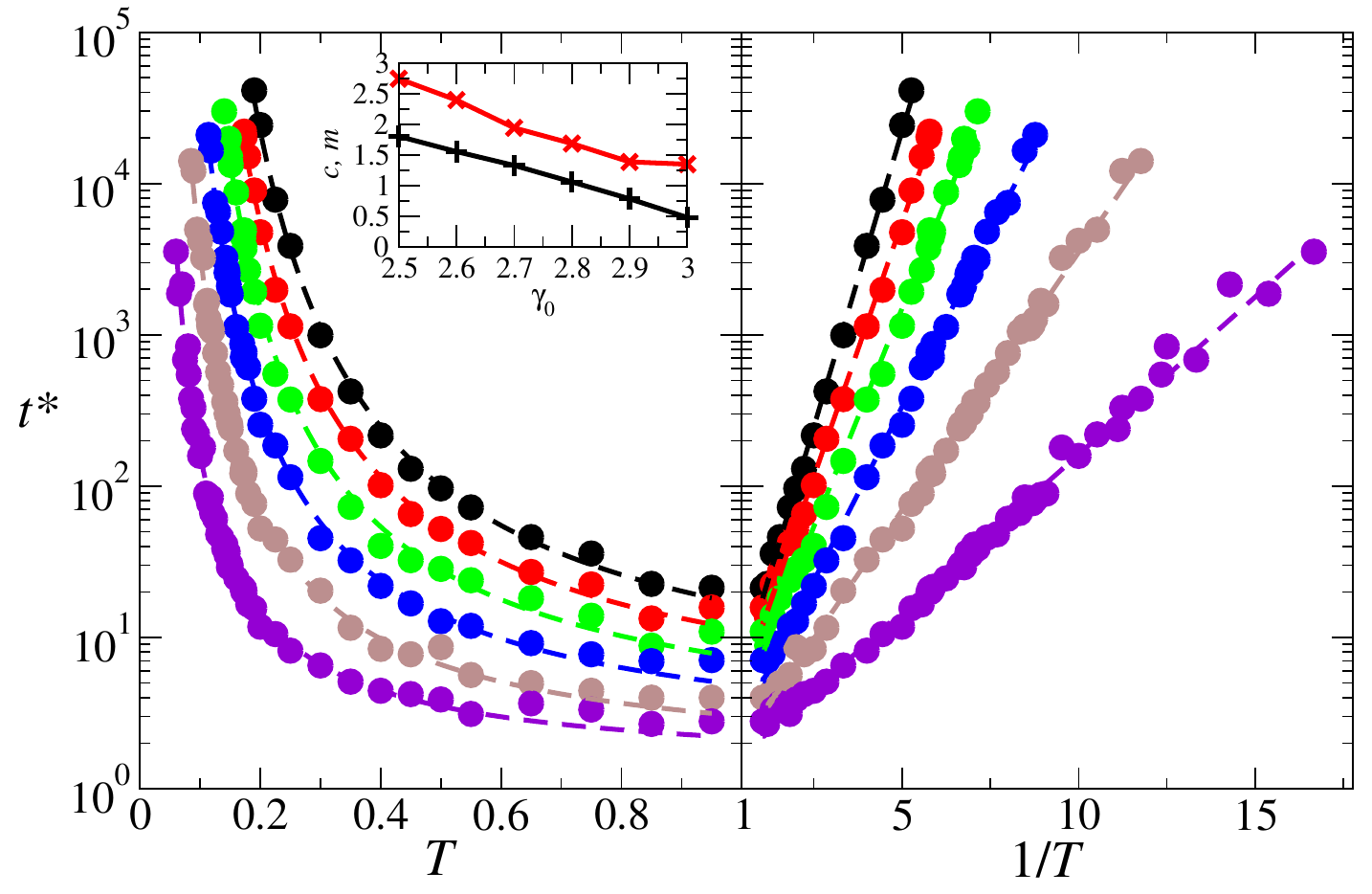}
  \caption{{\bf Soft glassy rheology model.}  Time delay $t^*$ after the imposition of a step strain before shear banding and material failure arise. Imposed step strain amplitude $\gamma_0=2.5,2.6,2.7,2.8,2.9,3.0$ in data sets downwards from black to violet.  {\bf Left:} Plotted as a function of the working temperature $T$. {\bf Right:} Plotted as a function of $1/T$. Solid symbols: numerical data. Dashed lines: fits to the exponential form $t^*=c\exp\left(m/T\right)$. Inset shows how the fitting variables $m$ ($+$) and $c$ ($\times$) depend on $\gamma_0$.  Waiting time $\tw=10^8$. }
  \label{fig:SGR_tStar_vs_x}
\end{figure}

So far, we have explored the phenomenon of delayed material failure by varying the strain amplitude $\gamma_0$, at a fixed level of sample annealing prior to shear. In Fig.~\ref{fig:SGRmodel_time_new} (right) we instead vary the level of sample annealing, as quantified by the time $\tw$ for which the system is allowed to age undisturbed before the strain is imposed. The same scenario pertains as in Fig.~\ref{fig:SGRmodel_time_new} (left). Plotting the delay time $t^*$ versus $\tw$ for different values of $\gamma_0$ in Fig.~\ref{fig:SGR_tStar_vs_tw_and_gamma0} (top right), we find the delay to increase apparently indefinitely with increasing sample age.

Within the SGR model constrained to have a homogeneous strain field, it can be shown exactly that the timescale for stress relaxation after a step strain scales  as $\tw\exp(-\gamma^2_0/2x)$, in the glass phase $T<T_{\rm g}$~\cite{fielding2000aging1}. To explore whether the delay time for catastrophic material failure  scales likewise, we rescale the ordinate of each curve  $t^*(\gamma_0)$  in Fig.~\ref{fig:SGR_tStar_vs_tw_and_gamma0} (top left) by  $\tw$. Excellent curve collapse confirms the basic scaling $t^*\sim \tw$ (bottom left) . We furthermore rescale the ordinate of each of the curves $t^*(\tw)$ in the top right panel by $\exp(-\alpha\gamma_0^2/2x)$ with $\alpha\approx 1.2$, finding good collapse for large $\tw$. 

Having explored the dependence of the delay time $t^*$ on the strain amplitude $\gamma_0$ and annealing level $\tw$, we consider finally its dependence on the working temperature $T$. In Fig.~\ref{fig:SGR_tStar_vs_x}  (left), we plot $t^*$ as a function of $T$ for several $\gamma_0$, at fixed $\tw$. As can be seen, $t^*$ increases dramatically with decreasing working temperature $T$. Re-scaling the horizontal axis $T\to 1/T$ in the right panel of Fig.~\ref{fig:SGR_tStar_vs_x} demonstrates a Boltzmann dependence, $t^*\sim \exp(m/T)$, with $m$ a constant, as may be expected intuitively.

\begin{figure}[!t]
  \includegraphics[width=8.5cm]{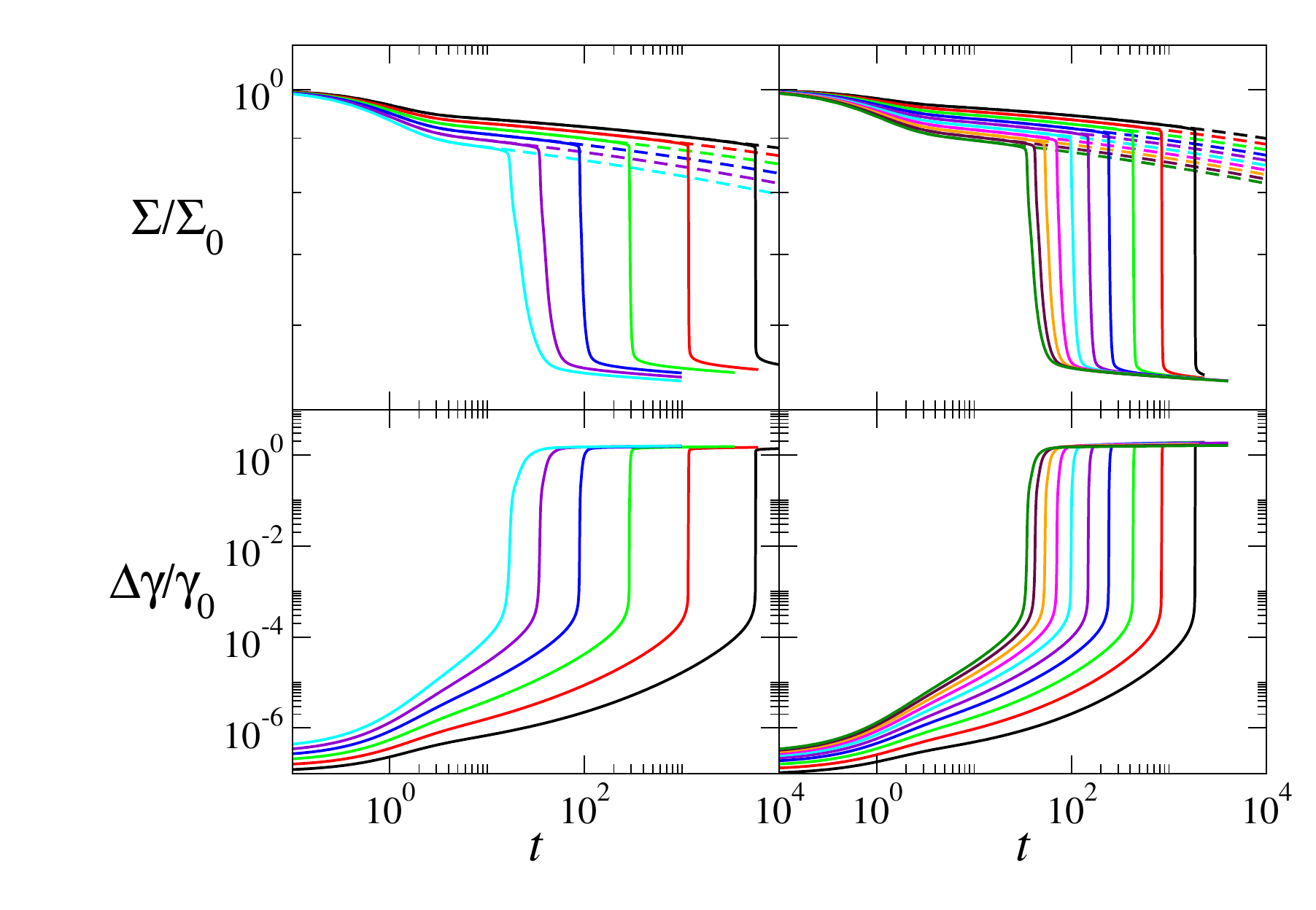}
  \caption{{\bf Elastoplastic model (1D).}   {\bf Top:} Stress
  decay versus time $t$ since the
  imposition of a step shear strain. Dashed curves show  calculations in which
  the strain field is artificially constrained to remain homogeneous,
  and solid in which it is allowed to become heterogeneous. {\bf Bottom:} Corresponding degree of strain heterogeneity $\Delta\gamma$. {\bf Left:} Strain amplitude $\gamma_0=1.16, 1.17\cdots 1.21$ in curve sets right to left, pre-quench temperature $T_0=0.02$.
  {\bf Right:} Strain amplitude $\gamma_0=1.2$, pre-quench temperatures $T_0=0.011, 0.012\cdots 0.020$ in curve sets right to left.
  Working temperature $T=0.01$. }
  \label{fig:EPM_time}
\end{figure}
\begin{figure}[!t]
\includegraphics[width=0.9\columnwidth]{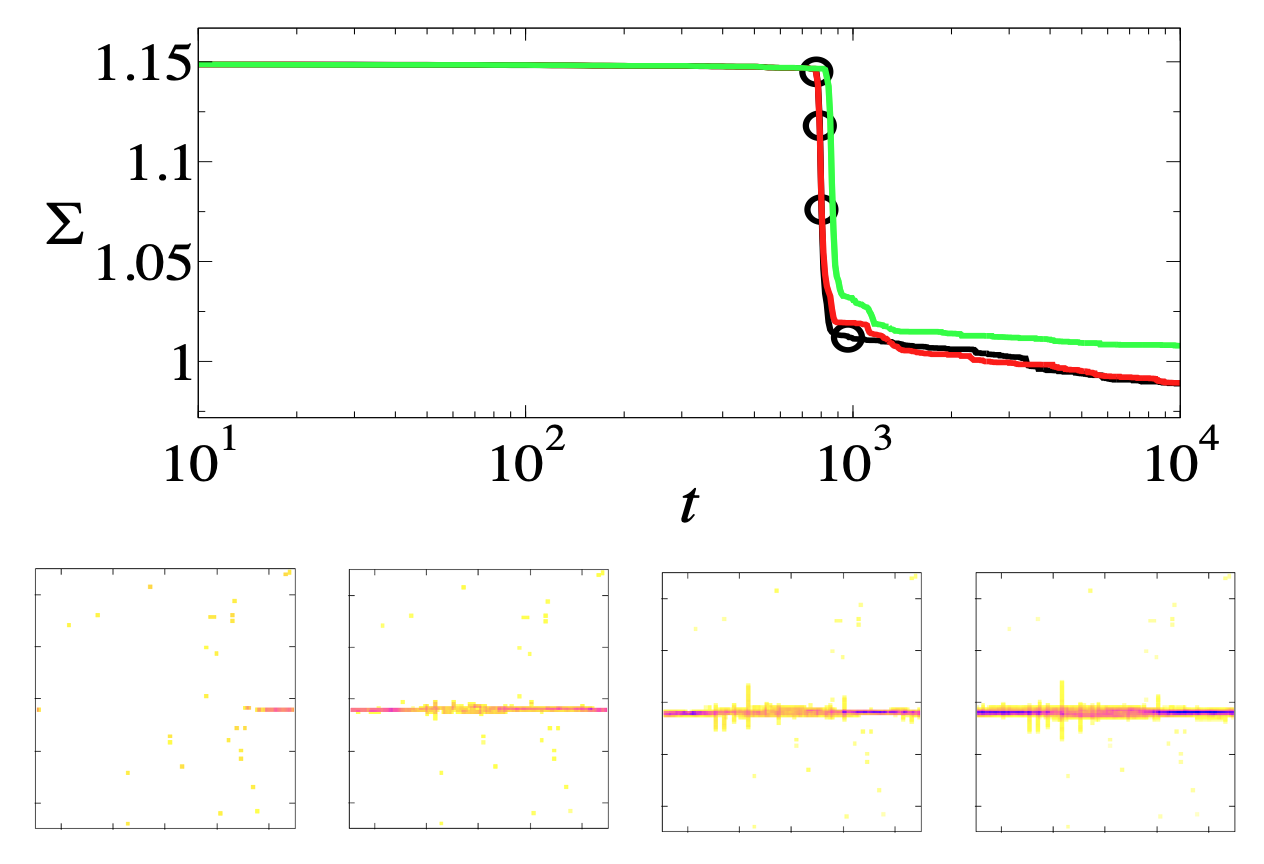}
  \caption{{\bf Elastoplastic model (2D).} Response to a shear strain of amplitude $\gamma_0=1.15$, linearly ramped at strain rate $\gdot=\gamma_0/t_{\rm ramp}$ over times $-t_{\rm ramp}<t<0$. System size $N^2$ with $N=128$. $T=T_0=0.01$.  {\bf Top:} stress decay for $t_{\rm ramp}=0.0,1.0,10.0$ (black, red, green). {\bf Bottom:} state snapshots showing accumulated number of local plastic yielding events at the times indicated by circles left to right in the top panel. 
}
  \label{fig:EPM_2D}
\end{figure}

\subsection*{Elastoplastic model}

The same physics also emerges in the 1D EPM. The top left panel of Fig.~\ref{fig:EPM_time} shows the stress decay following the imposition of a step shear strain for several different  strain amplitudes $\gamma_0$, at fixed working temperature $T$ and annealing temperature $T_0$. As in SGR, when the strain is constrained to remain homogeneous, the stress relaxes gradually. In contrast, when the strain is allowed to become heterogeneous, the stress suddenly drops precipitously after a delay time $t^*$ that increases steeply with decreasing $\gamma_0$. Corresponding to this is a sudden rise in the level of strain heterogeneity (bottom left panel), signifying a dramatic shear banding instability  and catastrophic material failure. The right panels of Fig.~\ref{fig:EPM_time} likewise demonstrate a strongly increasing failure time with increasing sample annealing prior to shear (lower $T_0$).

The same physics is also seen 
in 2D. The top panel of Fig.~\ref{fig:EPM_2D} shows the stress decay  versus time $t$ following the imposition of a strain in the 2D EPM~\cite{nicolas2018deformation,pollard2022yielding}. The state snapshots underneath show the accumulated number of local plastic events across the 2D lattice, at the times shown by circles in the top panel. Highly delayed material failure  is again evident, characterised by a precipitous stress drop and the formation of a shear band.

Throughout, we have modelled the strain as imposed instantaneously at time $t=0$. In  practice, finite inertia necessitates 
ramping the strain over a short but non-zero time interval $t_{\rm ramp}$. Fig.~\ref{fig:EPM_2D} (top) shows the physics we report to be robust to reasonable variations in $t_{\rm ramp}$.

\subsection*{Fluidity model}

So far, we have demonstrated the phenomenon of ultra-delayed material failure within the mesoscopic soft glassy rheology model in 1D and within a mesoscopic thermal elastoplastic model in 1D and 2D. Finally, we now confirm that the same basic physics arises in a simpler model still: our continuum fluidity model in 1D~\cite{moorcroft2011age}.

\begin{figure}[!t]
  \includegraphics[width=8.5cm]{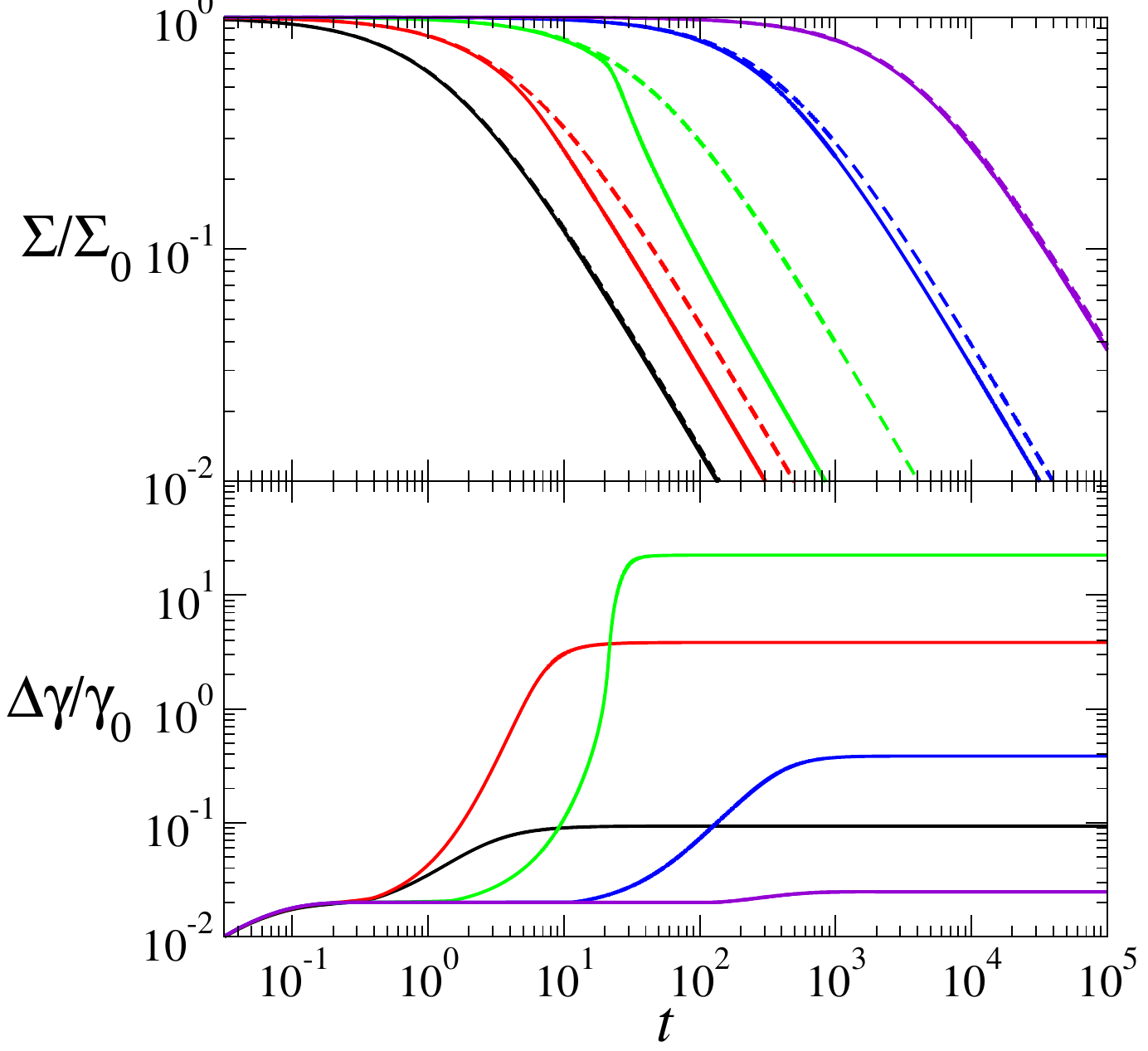} 
  \caption{{\bf Fluidity model} response  to a step strain of amplitude $\gamma_0=5.5$ applied at waiting times  $\tw=10^4,10^5,10^6,10^7,10^8$ in curve sets left to right. {\bf Top:} Stress
  decay as a function of time since the
  imposition of the strain.   Dashed lines show the results of calculations in which
  the strain field is artificially constrained to remain homogeneous,
  and solid lines in which it is allowed to become heterogeneous.  {\bf Bottom:} Corresponding degree of strain heterogeneity $\Delta\gamma$ across the
  sample, normalised by the imposed strain $\gamma_0$, with curve colours matching those in the top panel.
  \label{fig:fluidityModel_time}}
\end{figure}

Fig.~\ref{fig:fluidityModel_time} shows the response of the fluidity model to a step strain of a fixed amplitude $\gamma_0$ for several different levels of sample annealing prior to shear, as characterised by the sample age $\tw$. The top panel shows the normalised stress decay, and the bottom panel shows the corresponding degree of strain heterogeneity across the sample. As before, a strongly delayed shear banding instability is observed, accompanied by the formation of strain bands.

Despite the fact that the continuum fluidity model predicts the same basic phenomenon of delayed yielding as the mesoscopic models, one obvious difference between the SGR model and fluidity model is that the regime of significant banding instability is confined in the fluidity model  to a window of values of $\tw$ for any fixed $\gamma_0$: the instability is lost by taking a large enough initial sample age $\tw$, whereas it persists even as $\tw\to\infty$ at fixed $\gamma_0$ in the SGR model. We have no explanation for this difference. Because
the SGR model is a more sophisticated model that has been shown to
capture the rheology of yield stress fluids in numerous experimental
protocols, we suggest it to provide the better description.

\section*{Discussion}

We have predicted that catastrophic material failure can
arise at extremely long delay times after the imposition of a shear strain in amorphous materials, via a dramatic strain localisation instability, accompanied by a precipitous stress drop. In demonstrating the same basic physics within three constitutive models (SGR, EPM and fluidity), two methods of sample preparation prior to shear (ageing and temperature quenching), two methods of seeding a small initial heterogeneity, and in both one and two spatial dimensions, we suggest it may be generic across amorphous  materials.  

A particularly remarkable finding is that material failure can arise at extremely long delay times after the
initial strain imposition, i.e., long after the material last suffered
any deformation, with a delay time $t^*$ that increases dramatically with decreasing imposed strain amplitude $\gamma_0$, decreasing working temperature $T$, and increasing sample annealing prior to shear.  

\smf{We suggest the basic mechanism of the instability to be as follows. Imagine an initially near uniform sample, but with a streamline (in 1D) or localised region (in 2D or 3D) in which slightly more plastic stress relaxation arises, slightly fluidising that region relative to the rest of the sample.  Force balance must then be regained after that plastic relaxation. In 1D this results in the streamline in question straining slightly further forward still. In 2D or 3D, the stress propagation to neighbouring regions is determined by the Eshelby stress propagator~\cite{picard2004elastic,eshelby1959elastic}. This stress propagation then slightly predisposes nearby regions themselves plastically to relax. This knock-on effect represents a positive feedback loop, leading to a runaway instability of plastic yielding.}

\smf{In essence,  therefore, the basic physics  depends only on a combination of (i) plastic stress relaxation  activated on slow timescales, as may via arise via local rearrangements in particulate systems such as  dense colloidal suspensions, dense emulsions or metallic glasses, or bond breakage in network materials, with (ii) a follow-on propagation of stress to nearby regions from the site of the plastic stress loss. In view of this simple mechanism, the phenomenon reported here is likely to apply widely across many amorphous materials, independent of their detailed constituent substructure.}

 \smf{In future work, it would be interesting to explore any possible connection between the phenomenon reported here and the physics uncovered in Ref.~\cite{derlet2018thermally}, which performed molecular simulations of a Lennard-Jones material sheared first to a given strain at zero temperature, with the temperature then subsequently ramped up to just below the glass transition temperature. During this upward temperature ramp, and significant stress relaxation was observed, associated with a system-spanning shear event.}
 
A shear banding instability after a rapid shear strain has been observed
previously in polymer melts~\cite{boukany2009step,fang2011shear1},
although after a short delay time of just a few seconds (consistent
with the absence of long-term memory in those ergodic fluids), and
having its origin in a non-monotonic relationship between stress and
strain during the initial rapid straining
process~\cite{agimelen2013apparent2,moorcroft2014shear}. No such
non-monotonicity exists in any model explored here for an infinite
rate of strain imposition, suggesting a fundamentally different
instability mechanism from the one reported here in amorphous and soft glassy materials. 

We
hope this work will stimulate experiments and particle simulations aimed
at observing this instability.

Acknowledgments --- This project has received funding from the European Research Council (ERC) under the European Union's Horizon 2020 research and innovation programme (grant agreement No. 885146). We thank Andrew Clarke for discussions and SLB (Schlumberger Cambridge Research Ltd.) for funding.

%

\end{document}